\documentclass[11pt]{article}
\usepackage{moriond,epsfig}

\def\be{\begin{equation}}
\def\ee{\end{equation}}
\def\bea{\begin{eqnarray}}
\def\eea{\end{eqnarray}}

\begin{document}
\vspace*{4cm}
\title{NOVEL OPPORTUNITIES FOR EW BREAKING\\
 FROM LOW-SCALE SUSY BREAKING}

\author{ J.R. ESPINOSA}

\address{IFT-UAM/CSIC, Madrid, Spain}

\maketitle\abstracts{
In supersymmetric scenarios with a low scale of SUSY breaking
[$\sqrt{F}=O(TeV)$] the conventional MSSM Higgs sector can be 
substantially
modified, mainly because the Higgs potential contains additional effective
quartic terms. The Higgs spectrum can be dramatically changed, and the
lightest state can be much heavier than in usual SUSY scenarios. Novel
opportunities for electroweak breaking arise, and the electroweak scale
may be obtained in a less fine-tuned way for wide ranges of $\tan\beta$ 
and
the Higgs mass, offering a possible solution to the MSSM fine-tuning
problem.}

\def\meff{\langle m_{\nu} \rangle}

\newcommand{\ed}{\end{document}}
\newcommand{\bne}{\hbox{$\bar \nu_e$ }}
\newcommand{\nue}{\hbox{$\nu_e$ }} 
\newcommand{\nm}{\hbox{$\nu_\mu$ }} 
\newcommand{\bnm}{\hbox{$\bar \nu_\mu$ }}
\newcommand{\nt}{\hbox{$\nu_\tau$ }} 

\def\ltap{\raisebox{-.4ex}{\rlap{$\sim$}} \raisebox{.4ex}{$<$}}
\def\gtap{\raisebox{-.4ex}{\rlap{$\sim$}} \raisebox{.4ex}{$>$}}
\def\lsim{\raise0.3ex\hbox{$\;<$\kern-0.75em\raise-1.1ex\hbox{$\sim\;$}}}
\def\gsim{\raise0.3ex\hbox{$\;>$\kern-0.75em\raise-1.1ex\hbox{$\sim\;$}}}

\newcommand{\CL}   {C.L.}
\newcommand{\dof}  {d.o.f.}
\newcommand{\gof}{g.o.f.}
\newcommand{\EtAl} {{\it et al.\/}}
\newcommand{\eVq}  {\rm{eV}^2}
\newcommand{\Sol}  {\mathsc{sol}}
\newcommand{\Atm}  {\mathsc{atm}}
\newcommand{\JSQ}  {{Just-So$^2$}}

\newcommand{\Sbl}  {\mathsc{sbl}}
\newcommand{\Kl}  {\mathsc{KamLAND}}
\newcommand{\Nev}  {\mathsc{nev}}
\newcommand{\Lsnd} {\mathsc{lsnd}}
\newcommand{\Chooz}{\mathsc{chooz}}

\newcommand{\thl}{\theta_\Lsnd}
\newcommand{\sql}{\sin^2 2\thl}

\newcommand{\Dcq}  {\Delta\chi^2}
\newcommand{\Dms}  {\Delta m^2_\Sol}
\newcommand{\Dma}  {\Delta m^2_\Atm}
\newcommand{\Dml}{\Delta m^2_\Lsnd}
\newcommand{\Eps}  {\varepsilon}
\newcommand{\Epp}  {\varepsilon'}

\newcommand{\snocc}{SNO$_\mathrm{CC}^\mathrm{rate}$ }
\newcommand{\snotot}{SNO$_\mathrm{CC,NC}^\mathrm{SP,DN}$ }

\newcommand{\pnu}[1] {\overset{\smash{\scriptscriptstyle (-)}}{\nu}_{\hskip-3pt #1}}
\newcommand{\eps}{\epsilon_{13}}
\newcommand{\si}{S_{13}}
\newcommand{\SUSY}{\makebox[1.15cm][l]{$\line(4,1){30}$\hspace{-.95cm}{SUSY}}}
\newcommand{\tinySUSY}{\makebox[0.85cm][l]{$\line(4,1){19}$\hspace{-0.77cm}
{\tiny{SUSY}}}}

\def\simlt{\stackrel{<}{{}_\sim}}
\def\simgt{\stackrel{>}{{}_\sim}}
\def \znbb {0\nu\beta\beta}
\def \nbb {$\beta\beta_{0\nu}$ }

\def \tnbb {2\nu\beta\beta}
\def\meff{\langle m_{\nu} \rangle}
\def\rp{$R_p \hspace{-1em}/\;\:$}

\def\[{\left [}
\def\]{\right ]}
\def\({\left (}
\def\){\right )}

\section{Low-scale SUSY breaking}
\label{sec:1}

The simplest effective description of SUSY breaking (\SUSY) assumes that 
the theory contains a \SUSY sector not coupled directly to the observable 
sector, but rather through a heavier "messenger" sector (characterized by 
a mass scale $M$) coupled to both. Let us parameterize the \SUSY sector 
simply by a singlet chiral field $T$, whose $F$-term takes a non 
zero vacuum expectation value (vev) thereby breaking SUSY. After 
integrating out the heavy physics associated to the messenger sector, the 
low-energy theory for the observable fields $\varphi$ will contain 
non-renormalizable couplings between $T$ and $\varphi$. These operators 
give rise to soft terms (such as scalar soft masses), but
also hard terms (such as quartic scalar couplings):
\be
\label{mlambda}
\int d^4\theta {1\over M^2} |T|^2|\varphi|^2\Rightarrow m_{\rm soft}^2\sim 
{F^2\over M^2}\ ;\;\;\;\; 
\int d^4\theta {1\over M^4} |T|^2|\varphi|^4\Rightarrow 
\lambda_{\tinySUSY}\sim
{F^2\over M^4}\sim {m_{\rm soft}^2\over M^2}\ .
\ee
Phenomenology requires $m_{\rm soft} = {\cal O}(1\, {\rm TeV})$, but 
this leaves undetermined the scale $\sqrt{F}$ of \SUSY. We can consider 
two extreme possibilities: {\it 1)} In models with \SUSY mediated by 
gravity $M\sim 10^{18}$ GeV implies $\sqrt{F}\sim 10^{11}$ GeV. In such 
cases 
the large hierarchy $M\gg \sqrt{F}\gg m_{\rm soft}$ makes unobservable the 
$\lambda_{\tinySUSY}$ contributions. {\it 2)} If $M$ is not far from the 
TeV scale, then $\sqrt{F}$ would be of the same order. We are interested 
in such low-scale \SUSY scenarios\cite{hard,Brignole,Polonsky,BCEN}, in 
particular to explore the implications of the hard terms in
eq.~(\ref{mlambda}) for electroweak symmetry breaking. 

In ref.\cite{BCEN} we performed a model independent analysis of such 
effects using this effective theory approach. The Higgs potential of 
the resulting low-energy effective theory is a generic two Higgs doublet 
model (THDM). In terms of the invariants ${\cal I}_1=|H_1|^2$,  ${\cal 
I}_2=|H_2|^2$, ${\cal I}_{12}=H_1\cdot H_2$: 
\bea
V&=&m_1^2 {\cal I}_1 + m_2^2 {\cal I}_2 + (m_3^2 {\cal I}_{12}+{\mathrm 
h.c.})\nonumber\\
&+&{1\over 2}\lambda_1 {\cal I}_1^2+{1\over 2}\lambda_2 
{\cal I}_2^2
+\lambda_3 {\cal I}_1{\cal I}_2+\lambda_4|{\cal I}_{12}|^2+
\left[{1\over 2}\lambda_5{\cal I}_{12}^2+(\lambda_6{\cal 
I}_1+\lambda_7{\cal I}_2){\cal I}_{12} +{\mathrm
h.c.}\right]
\ ,
\label{THDMpot}
\eea
with $T$-dependent coefficients. (If the $T$ field is heavy enough 
it can be integrated out.)
The \SUSY contributions to the Higgs quartic 
couplings are no longer negligible and can be easily 
larger than the ordinary MSSM values, given by gauge couplings, and this
has a deep impact on
the pattern of EW breaking. Some of the main differences with
respect to the MSSM are the following\cite{BCEN}. {\it a)} There is no need of
radiative corrections to trigger EW breaking. It occurs already at
tree-level (which is just fine since the effects of RG running are
small for a low cut-off scale $M$). {\it b)} EW breaking occurs naturally
only in the Higgs sector, as desired. {\it c)} The parameter space that
gives a proper EW breaking is larger (the constraints from unbounded from
below directions can be absent). {\it d)} The Higgs spectrum and
properties can be very different due to very different quartic couplings.  
In particular, the MSSM upper bound on the mass of the lightest Higgs
field no longer applies. In fact, all Higgses might be significantly
heavier than the $Z^0$. {\it e)} There is an extra degree of freedom in 
the scalar sector, coming from the complex singlet $T$. {\it f)} The Higgs 
quartic couplings $\lambda_{5},\lambda_{6},\lambda_{7}$ can be complex and 
introduce new possible sources of $CP$ violation already at tree level.
{\it g)} As we will see, the changes of the Higgs potential allow for a 
drastic reduction 
of the fine tuning required for the EW breaking.

\section{A concrete model}
\label{sec:2}

\begin{figure}[t]
\centerline{
\psfig{figure=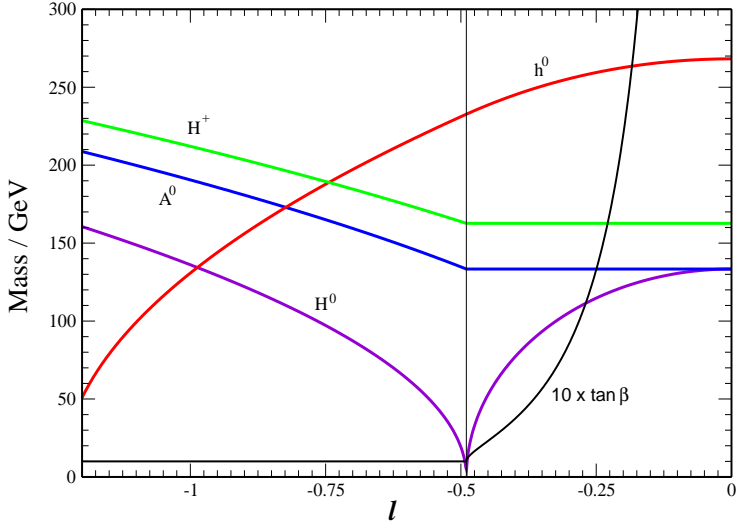,height=8cm,width=7cm,angle=-90,bbllx=3.cm,%
bblly=4.cm,bburx=20.cm,bbury=24.cm}}
\caption{\footnotesize
Higgs spectrum of the example model as a function of $\ell$. Also shown 
is $\tan\beta$ (scaled by a factor 10).}
\label{specA}
\end{figure}

Let us now present a concrete example\cite{BCEN}. 
The superpotential is given by 
(with the notation introduced above, plus ${\cal T}=|T|^2$)
\be 
W =\Lambda_S^2 T + \mu {\cal I}_{12} + {\ell\over 2M}{\cal I}_{12}^2 \ ,
\ee
and the K\"ahler potential is
\be 
K  = {\cal T}+{\cal I}_1+{\cal I}_2 - 
 {\alpha_t \over 4 M^2} {\cal T}^2 + 
{\alpha_1 \over M^2}{\cal T} ({\cal I}_1+{\cal I}_2) + 
{e_1 \over 2M^2}\left({\cal I}_1^2+{\cal I}_2^2\right)\ .
\ee
(All parameters are real with $\alpha_t>0$.) Here $\Lambda_S$ 
is the \SUSY scale and $M$ the `messenger' scale (see previous section).
The typical soft masses are $\sim \tilde{m}\equiv \Lambda_S^2/M$. In 
particular, the mass of the scalar component of $T$ is 
${\cal{O}}(\tilde{m})$ and, after integrating this field out, the 
effective potential for $H_1$ and $H_2$ is a 2HDM  with
very particular Higgs mass terms: $m_1^2=m_2^2=\mu^2-\alpha_1 
\tilde{m}^2$, $m_3^2=0$ and Higgs quartic
couplings like those of the MSSM plus contributions of order $\mu/M$ and
$\tilde{m}^2/M^2$:
\bea
\label{quarticc}
\lambda_1=\lambda_2&=&{1\over 4}(g^2+g_Y^2)+2\alpha_1^2{\tilde{m}^2\over
M^2}\ ,\nonumber\\ \lambda_3&=&{1\over 4}(g^2-g_Y^2)+{2\over
M^2}(\alpha_1^2\tilde{m}^2-e_1\mu^2)\ ,\nonumber\\  \lambda_4&=&-{1\over
2}g^2-2\left(e_1+2{\alpha_1^2\over \alpha_t}\right){\mu^2\over M^2}\
,\nonumber\\  \lambda_5&=&0\ ,\nonumber\\
\lambda_6=\lambda_7&=&{\ell\mu\over M}\ .
\eea
The explicit expressions for the spectrum of Higgs masses 
can be found in \cite{BCEN}. Here we plot the masses as a function of 
the superpotential parameter $\ell$ 
in fig.~\ref{specA}, together with $\tan\beta\equiv \langle 
H_2\rangle/\langle H_1\rangle$, which is 1 if $\ell$ is 
below some critical value. The differences with the MSSM are apparent.

\section{The fine tuning problem of the MSSM}
\label{sec:3}

The MSSM Higgs sector is a very constrained THDM. The potential 
$V^{\rm MSSM}$ is of the 
form (\ref{THDMpot}) with $m_{1,2}^2=\mu^2 + m_{H_{1,2}}^2$ and 
$m_3^2=B\mu$, where $m_{H_{i}}^2$ and $B$ are soft masses and $\mu$ is the
Higgs mass term in the superpotential, $W\supset \mu H_1\cdot  H_2$. The 
quartic Higgs couplings are simply
\be
\label{VMSSM}
V_4^{\rm MSSM}=\frac{1}{8}(g^2+g_Y^2) (|H^0_1|^2-|H^0_2|^2)^2\ .
\ee
Minimization of $V^{\rm MSSM}$ leads to a vev $v^2\equiv 2(\langle H^0_1 
\rangle^2 + \langle H^0_2\rangle^2)$ and thus to a mass for the $Z^0$ 
gauge boson, $M_Z^2=\frac{1}{4}(g^2+g_Y^2) v^2$.

The size of $v$ is determined by the parameters of $V^{\rm MSSM}$, in
particular $m_i^2$, which depend on the initial parameters, $p_\alpha$
[for the MSSM these are the soft scalar masses $m$, the $\mu-$parameter, 
the gaugino masses $M$ and the trilinear soft terms $A$ at the
initial (high energy) scale]. Therefore, $v^2=v^2(p_1, p_2, \cdots)$.
As an example, for the case of large $\tan\beta$ and taking the initial 
scale to be $M_{GUT}$ one gets
\be
M_Z^2\simeq -2.02 \mu_0^2 + 3.57 M_0^2 + 0.07 m_0^2 + 0.22 A_0^2 + 0.75 
A_0 M_0\ ,
\label{topdown}
\ee
where the subindex $0$ indicates values evaluated at $M_{GUT}$ and we have 
assumed universality. Getting the right value for $M_Z$ (or $v$) in a 
natural way requires that the $p_\alpha$'s are not too large, 
which in turn requires that the masses of superpartners should be  
$\simlt$ few hundred
GeV. Actually, the available experimental data already imply that the 
ordinary MSSM is significantly fine 
tuned\cite{BG,dCC,Poko,Anderson,Paolo,totum}.
Consider for instance the lower bound on the lightest Higgs boson mass
\bea
\label{mhMSSM}
m_h^2\leq M_Z^2 \cos^2 2 \beta + {3 m_t^4 \over 2\pi^2 v^2}
\log{M_{\rm SUSY}^2\over m_t^2} + ...  \eea
where $m_t$ is the (running) top mass ($\simeq 171$ GeV for $M_t=178$ 
GeV). Since the experimental
lower bound, $(m_h)_{\rm exp}\geq 115$ GeV, 
exceeds the tree-level contribution, the radiative
corrections must be responsible for the difference, and this translates
into a lower bound on $M_{\rm SUSY}$:
\be
\label{MSUSYMSSM}
M_{\rm SUSY}\;\simgt\;  e^{-2.1\cos^2 2\beta}
 e^{\left({m_h}/{62\ {\rm GeV}}\right)^2} m_t\;\simgt\;   3.8\ m_t\ ,
\ee
where the last figure corresponds to $m_h= 115$ GeV and large $\tan
\beta$, which is the most favorable case for the fine tuning. 
The last equation implies sizeable soft terms, 
$\tilde m \simgt 2 m_t$, which in turn translates into large
fine-tunings.

We adopt as a measure of the fine tuning associated to $p_\alpha$ the 
quantity $\Delta_{p_\alpha}$ defined by \cite{BG}
\be
\label{BG}
{\delta v^2\over v^2} = 
\Delta_{p_\alpha}{\delta
p_\alpha\over p_\alpha}\ ,  
\ee
where $\delta v^2$ is the change induced in 
$v^2$ by a change $\delta p_\alpha$ in $p_\alpha$. 
Absence of fine tuning requires
that $\Delta_{p_\alpha}$ should not be larger than 
${\cal O}(10)$.\footnote{Roughly speaking
$\Delta^{-1}_{p_\alpha}$ measures the probability of a cancellation
among  terms of a given size to obtain a result which is
$\Delta_{p_\alpha}$ times  smaller. For discussions see 
\cite{dCC,Poko,Anderson,Paolo}.}

\begin{figure}[t]
\centerline{
\psfig{figure=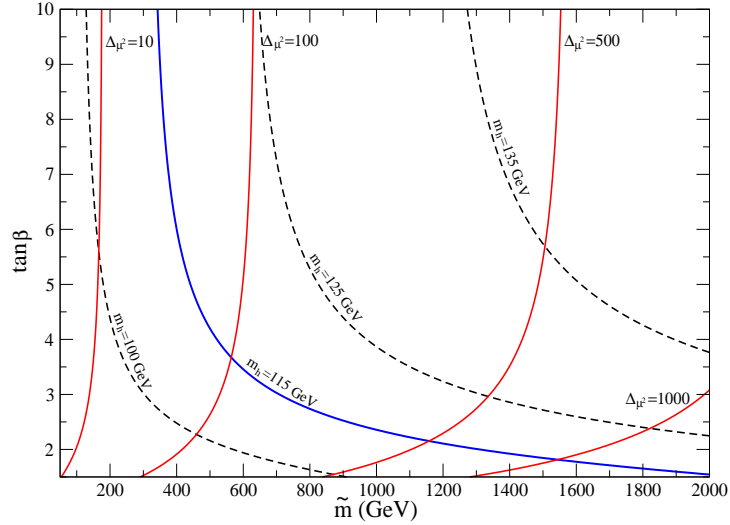,height=8cm,width=7cm,angle=-90,bbllx=3.cm,%
bblly=4.cm,bburx=20.cm,bbury=24.cm}}
\caption{\footnotesize
Fine tuning in the MSSM (measured by $\Delta_{\mu^2}$, solid lines) in the
$(\tilde{m},\tan\beta)$ plane. Dashed lines are contour lines of constant
Higgs mass.}
\label{ftMSSM}
\end{figure}

The fine-tuning problem of the MSSM is illustrated by fig.~(\ref{ftMSSM})
which has $\tilde{m}=M_0=m_0=A_0$. 
It shows how the fine-tuning $\Delta_{\mu^2}$
\footnote{$\mu^2$ is the parameter that usually 
requires the largest fine 
tuning since, due to the negative sign of its contribution in
eq.~(\ref{topdown}), it has to compensate the (globally
positive and large) remaining contributions.}
grows with 
increasing $\tilde{m}$ and decreasing $\tan\beta$. Insisting in 
$\Delta_{\mu^2}<10$ would require $m_h<105$ GeV and $\tilde{m}<175$ GeV. 
It also shows that for soft masses $\tilde{m}^2\sim a v^2$ the 
fine-tuning scales like $\Delta_{\mu^2}\sim 20 a$. That is, the MSSM 
fine-tuning is much larger than one would naively expect  
($\Delta_{\mu^2}\sim a$).

Why is this so? There are three main reasons [{\it 1)} to {\it 3)} below]. 
To better understand them let us write the Higgs potential along the 
breaking direction in the $H_1^0, H_2^0$ space. It is of SM-like form:
\be
\label{Vbeta}
V={1 \over 2} m^2 v^2 + {1 \over 4} \lambda v^4 \ ,
\ee
where $\lambda$ and $m^2$ are functions of the $p_\alpha$ parameters and 
$\tan\beta$, in particular $m^2 = 
c_\beta^2\, m_1^2(p_\alpha)+s_\beta^2\, m_2^2(p_\alpha)-s_{2\beta}\, 
m_3^2(p_\alpha)$.
Minimization of (\ref{Vbeta}) leads to
\be
\label{v2}
v^2={-m^2\over \lambda} \ .
\ee

{\it 1)} In the MSSM $\lambda$
turns out to be quite small:
\be
\label{lambdaMSSM}
\lambda_{\rm MSSM}={1 \over 8}(g^2+g_Y^2)\cos^2 2\beta\ \simeq \  {1
\over 15}\cos^2 2\beta\ .
\ee
This has the effect of amplifying whatever cancellations are taking place 
inside $m^2$ in (\ref{v2}). This implies a fine tuning $\simgt 15$ times 
larger than expected from naive dimensional considerations.
The previous $\lambda_{\rm MSSM}$ was evaluated at tree-level but
radiative corrections make $\lambda$ larger, thus reducing the fine
tuning\cite{dCC,Poko}.  Since $m_h^2 \sim 2\lambda v^2$, the ratio
$\lambda_{\rm tree}/\lambda_{\rm 1-loop}$ is basically the ratio
$(m_h^2)_{\rm tree}/(m_h^2)_{\rm 1-loop}$, so for large $\tan \beta$ the
previous factor 15 is reduced by a factor $M_Z^2/m_h^2$. 

{\it 2)} Although for a given size of the soft terms the radiative
corrections reduce the fine tuning, sizeable radiative corrections require
large soft terms, which in turn worsens the fine tuning.  A given increase
in $\tilde{m}^2$ reflects linearly in $m^2$ but only logarithmically in 
$\lambda$, so the fine
tuning usually gets worse. In fact, the Higgs bound (\ref{mhMSSM}) implies
that radiative corrections should be sizeable and therefore this restricts
the allowed parameter space to regions of large fine tuning.

{\it 3)} Thanks to supersymmetry, $m^2$ is not sensitive to large 
quadratic corrections in the MSSM but it receives sizeable logaritmic 
corrections $\delta m^2\propto (\tilde{m}^2/16\pi^2)\log (M_{GUT}^2/\tilde 
m^2)$, which can be interpreted as the effect of the RG running of $m^2$ 
from $M_{GUT}$ down to the electroweak scale.  
Typically, the large logarithms and the numerical factors
compensate the one-loop factor, so that the corrections are quite
large. This is the origin of the large coefficients that appear in 
formulae such as (\ref{topdown}) which also have an impact on the value of 
the fine-tuning.

\section{Fine tuning in models with low-scale SUSY breaking}
\label{sec:4}

It is remarkable that low-scale \SUSY models of the type discussed in the 
first two sections have all the necessary ingredients to ameliorate 
the shortcomes of the MSSM regarding the fine-tuning of EW breaking. 
Let us see this following the same numbering as above:

{\it 1)} As explained in Sect.~1, tree-level 
quartic Higgs couplings larger than in the MSSM are natural in scenarios 
in which the breaking of SUSY occurs at a 
low-scale\cite{hard,Brignole,Polonsky,BCEN} (not far from the TeV 
scale).\footnote{This can also happen in 
models with extra dimensions opening up not far from the electroweak scale 
\cite{Strumia}. Another way of increasing
$\lambda_{\rm tree}$ is to extend the gauge sector \cite{extG} or to
enlarge the Higgs sector \cite{extH}. The latter option has been studied
in \cite{NMSSM} (for the NMSSM) but this framework is less effective in
our opinion.} 

{\it 2)} If the quartic Higgs couplings are already large at tree level,
the LEP Higgs mass bound can be evaded without the need of large radiative 
corrections, and therefore regions of parameter space with lower soft 
masses are not excluded and in them the fine-tuning is naturally smaller.

{\it 3)} In models with a low SUSY breaking scale RG effects
are expected to play no significant role since the cut-off scale is much 
closer to the electroweak scale.

\begin{figure}[t]
\centerline{
\psfig{figure=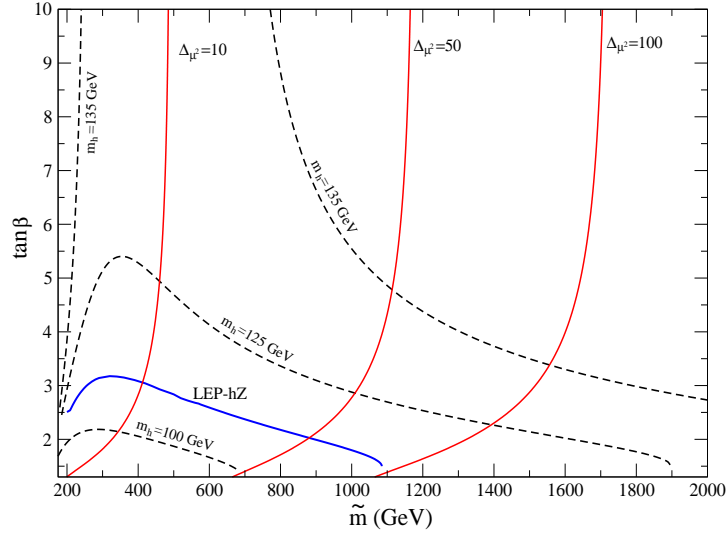,height=8cm,width=7cm,angle=-90,bbllx=3.cm,%
bblly=4.cm,bburx=20.cm,bbury=24.cm}}
\caption{\footnotesize
Fine tuning in the $(\tilde{m},\tan\beta)$ plane
in the low-scale SUSY breaking
scenario of Sect.~2.
Dashed lines are contour lines of constant $m_h$. The LEP bound is also shown.
}
\label{ftlow}
\end{figure}

All these three improvements can cooperate to make EW breaking much more 
natural than in the MSSM. We show this in fig.~(\ref{ftlow}) for the model 
of Sect.~2. The reader should keep in mind that this model was not 
constructed with the goal of reducing fine-tuning in mind.
The corresponding expression for 
$\Delta_{\mu^2}$, as evaluated from eq.(\ref{BG}), is 
\be
\Delta_{\mu^2}=-{\mu^2\over \lambda v^2}\left[1+v^2\left(
{ls_{2\beta}\over 2\mu M}-{1\over M^2}
(e_1+{\alpha_1^2\over \alpha_t})
s_{2\beta}^2
\right)\right]\ .
\label{newfine}
\ee
%

In fig.~\ref{ftlow} we see an overall decrease in the size of 
fine-tuning of one 
order of magnitude with respect to the MSSM.  Also, restricting the fine 
tuning to be less than 10 does not
impose an upper bound on the Higgs masses, in contrast with the MSSM case.
As a result, the LEP bounds do not imply a large fine tuning.
 In any case, for $\Delta_{\mu^2}\leq 10$ we do find an upper bound
$\tilde{m}\simlt 500$ GeV, so that LHC would either find superpartners or
revive an (LHC) fine tuning problem for these scenarios (although the
problem would be much softer than in the MSSM).

\vspace{5pt}

\noindent
I would like to thank Andrea Brignole, Alberto Casas,
Irene Hidalgo and Ignacio Navarro for a most enjoyable collaboration.
This work is supported in part by the Spanish Ministry of Science and 
Technology through a MCYT project (FPA2001-1806).

\end{document}